\documentclass[prb,twocolumn,superscriptaddress]{revtex4}

\usepackage{graphicx}% Include figure files
\usepackage{amssymb,amsmath}
\usepackage{dcolumn}% Align table columns on decimal point
\usepackage{bm}% bold math
\usepackage{color}
\begin{document}

\title{Hall effect for indirect excitons in an inhomogeneous magnetic field}

\author{K. B. Arnardottir}
\affiliation{Science Institute, University of Iceland, Dunhagi-3,
IS-107, Reykjavik, Iceland}
\affiliation{Fysikum, Stockholms Universitet, S-106 91 Stockholm, Sweden}

\author{O. Kyriienko}
\affiliation{Science Institute, University of Iceland, Dunhagi-3,
IS-107, Reykjavik, Iceland}
\affiliation{Division of Physics and Applied Physics, Nanyang Technological University 637371, Singapore}

\author{I. A. Shelykh}
\affiliation{Science Institute, University of Iceland, Dunhagi-3,
IS-107, Reykjavik, Iceland}
\affiliation{Division of Physics and Applied Physics, Nanyang Technological University 637371, Singapore}

\date{\today}% It is always \today, today,
             %  but any date may be explicitly specified

\begin{abstract}
We study the effect of an inhomogeneous out-of-plane magnetic field on the behaviour of 2D spatially indirect excitons. Due to the difference of the magnetic field acting on electrons and holes the total Lorentz force affecting the center of mass motion of an indirect exciton appears. Consequently, an indirect exciton acquires an effective charge proportional to the gradient of the magnetic field. The appearance of the Lorentz force causes the Hall effect for neutral bosons which can be detected by measurement of the spatially inhomogeneous blueshift of the photoluminescence using counterflow experiment.
\end{abstract}

\maketitle

\section{Introduction}
The interaction of electrically charged particles with an external
magnetic field leads to a variety of phenomena in condensed matter
physics. The classical examples are integer and fractional quantum
Hall effects which have been studied extensively both experimentally
and theoretically in the last
decades.\cite{vonKlitzing,Tsui,Laughlin} In the domain of cold atoms
the magnetic field can also change the behavior of the system,
\textit{e.g.} driving the BEC-BCS transition by means of Feshbach
resonances.\cite{Feshbach}  However, as cold atoms are neutral
objects, the application of the magnetic field does not lead to an
appearance of the Lorentz force and Hall effect. Meanwhile, the
possibility of generation of an artificial magnetic field in the
atomic systems was proposed.\cite{Dalibard} This phenomenon is based
on the effect of the geometric (Berry) phase and requires
illumination of the sample by several laser beams tuned in resonance
with atomic transitions.\cite{Spielman} This has opened the way for
the observation of the analog of quantum Hall effect for neutral
cold bosons and fermions.

In solid state physics there exist electrically neutral bosonic
particles similar to atoms. These are excitons --
bounded electron-hole pairs. The impact of excitons onto optical and
transport properties of semiconductor materials have been studied
intensively,\cite{Knox} and the possibility of Bose-Einstein
condensation for excitons was consider theoretically long time
ago.\cite{Keldysh} However, an experimental observation of exciton
BEC still remains an open question.\cite{Moskalenko} A great step
forward was achieved by using the effect of strong exciton-photon
coupling in semiconductor microcavities.\cite{KavokinBook} Hybrid
light-matter quasiparticles called exciton-polaritons revealed
intriguing physical properties and the formation of a
macroscopically coherent state of polaritons was experimentally
reported.\cite{KasprzakNature} However, the question of how it is
connected with the standard BEC picture is still under
debate.\cite{ButovKavokin} Indeed, usually cavity polaritons have a
very short lifetime (not exceeding tens of picoseconds) which
prevents the possibility of full thermalization of the system.

\begin{figure}
\includegraphics[width=0.75\linewidth]{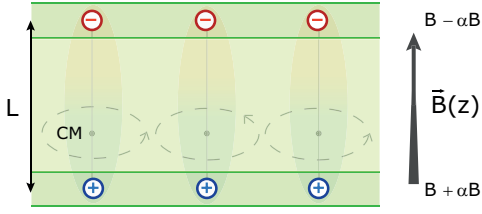}
\caption{ (color online). Sketch of the geometry of the system. Electrons and holes located in two QWs separated by the distance $L$ form spatially indirect excitons. The inhomogeneous in $z$ direction magnetic field $\mathbf{B}(z)$ affects the orbital motion of the center-of-mass (CM) of exciton.}
\label{Fig1}
\end{figure}

Other candidates proposed for the achievement of BEC in condensed
matter systems are indirect excitons, composite quasiparticles
consisting of electrons and holes located in spatially separated
quantum wells and bound together by Coulomb attraction (see sketch
in Fig. \ref{Fig1}). They obey bosonic statistics and can undergo
Bose-Einstein condensation at temperatures around
$1~K$.\cite{LozovikYudson} Spatially indirect excitons differ from
the usual 3D or 2D direct excitons by a much longer lifetime,
robustness with respect to the effects of disorder and much stronger
exciton-exciton interactions provided by dipole-dipole repulsion.
Consequently, much experimental efforts were devoted for observation
of emergence of spontaneous coherence of the indirect exciton gas
and macroscopically ordered exciton
state.\cite{Butov,Butov2002,ButovNature,Kyriienko,Liew}

Properties of an exciton subjected to magnetic field were studied
intensively\cite{Elliot,Hasegawa,Gorkov} and two qualitatively
different behaviors of exciton were found. In the case of small
magnetic field, where magnetic length is much bigger than exciton
Bohr radius, the center-of-mass motion of the exciton is not
affected, while its internal structure is modified.\cite{Gorkov} In
the opposite limit of strong magnetic fields an electron and a hole
from lowest Landau levels form \emph{magnetoexciton} with properties
different from the conventional exciton case.\cite{Lozovik2} Their
internal structure and center of mass motion are coupled, and being
exposed into crossed electric and magnetic field exhibit motion
perpendicular to both directions due to collinearity of the drift
velocities of the particles of opposite charge.
\cite{Lozovik2,Imamoglu,Paquet}

In this article we study the behavior of spatially indirect excitons
in the \emph{weak} magnetic field which is inhomogeneous in
out-of-plane $z$ direction. While in this limit a homogeneous
magnetic field only changes the binding energy of the exciton, the
inhomogeneous magnetic field acts differently on electrons and
holes and results in appearance of the Lorentz force acting on the
center-of-mass (CM) of a neutral exciton. Consequently, an exciton
acquires an effective ``charge'' and observation of an analog of the
Hall effect becomes possible. One should note that the following
mechanism is qualitatively different from the case of
magnetoexcitons described above.

\section{Exciton in an inhomogeneous magnetic field} We start by considering the quantum problem of a single indirect exciton in a weak external magnetic field.\cite{Elliot,Hasegawa,Gorkov,Edelstein} The generic quantum Hamiltonian of an electron and a hole in different quantum wells separated by distance $L$ with an external magnetic field pointed in $z$-direction perpendicular to the quantum well (QW) plane reads
\begin{align}\label{H0}
&\hat{H}=\frac{\hbar^{2}}{2m_{e}}\Big(-i\nabla_{e}+\frac{e}{c}\mathbf{A}_{e}(\mathbf{r}_{e})\Big)^{2}+\\&+\frac{\hbar^{2}}{2m_{h}}\Big(-i\nabla_{h}-\frac{e}{c}\mathbf{A}_{h}(\mathbf{r}_{h})\Big)^{2}-\frac{e^{2}}{\varepsilon\sqrt{(\mathbf{r}_{e}-\mathbf{r}_{h})^{2}+L^{2}}},
\notag
\end{align}
where $\mathbf{r}_{e,h}$ are the radius-vectors of 2D electrons and holes, $m_e$ and $m_h$ are their effective masses and $L$ is the separation between QWs. In the following consideration we restrict ourselves to the approximation of narrow QWs. Taking into account the finite width of the QW is straightforward but requires additional computational efforts.\cite{Lozovik} The external magnetic field acting on an electron and a hole is introduced into the Hamiltonian via the vector potential $\mathbf{A}(\mathbf{r})$. The magnetic field is considered to be inhomogeneous in $z$ direction and therefore the vector potentials for electrons and holes $\mathbf{A}_{e}(\mathbf{r}_{e}),~\mathbf{A}_{h}(\mathbf{r}_{h})$ are different.

The following Hamiltonian for an electron-hole pair with attraction can be rewritten in terms of CM coordinates $\mathbf{R}=\beta_{e}\mathbf{r}_{e}+\beta_{h}\mathbf{r}_{h}$ and relative motion coordinates $\mathbf{r}=\mathbf{r}_{e}-\mathbf{r}_{h}$, where $\beta_{e,h}=m_{e,h}/M$ and $M=m_{e}+m_{h}$. Thus, Eq. (\ref{H0}) can be recast as
\begin{align}\label{H1}
&\hat{H}=\frac{\hbar^{2}}{2m_{e}}\Big(-i\beta_{e}\nabla_{R}-i\nabla_{r}+\frac{e}{c}\mathbf{A}_{e}(\mathbf{R}+\beta_{h}\mathbf{r})\Big)^{2}+\\&+\frac{\hbar^{2}}{2m_{h}}\Big(-i\beta_{h}\nabla_{R}+i\nabla_{r}-\frac{e}{c}\mathbf{A}_{h}(\mathbf{R}-\beta_{e}\mathbf{r})\Big)^{2}-\frac{e^{2}}{\varepsilon\sqrt{r^{2}+L^{2}}},
\notag
\end{align}
where $\nabla_{R,r}$ are vector differential operators for CM and relative exciton motion coordinates, respectively.

We consider the case of a magnetic field $\mathbf{B}(z)$ that is inhomogeneous in $z$ direction. Within the approximation of narrow QWs this means that 2D electrons and holes in parallel layers feel different values of $B$. In the following derivation we use the symmetric gauge for the vector potentials $\mathbf{A}_i(\mathbf{r})=[\mathbf{B}_i\times \mathbf{r}]/2$ with values of the magnetic field taken different for the electrons and the holes, $B_{e}=B-\Delta B/2=B(1-\alpha)$ and $B_{h}=B+\Delta B/2=B(1+\alpha)$, where the parameter $\alpha$ describes the degree of the inhomogeneity of the magnetic field, $\alpha\approx(L/2B)dB/dz$.

It is convenient to rewrite the Hamiltonian (\ref{H0}) using the phase shift transformation of the wavefunction of the exciton similar to those used for the case of the homogeneous magnetic field \cite{Gorkov,Freire}
\begin{equation}
\Psi '(\mathbf{R},\mathbf{r})=\exp \Big(-i\frac{e}{2\hbar c}[\mathbf{B}\times \mathbf{r}]\cdot \mathbf{R}\Big)\Psi(\mathbf{R},\mathbf{r}).
\label{psi}
\end{equation}
The new Hamiltonian reads
\begin{widetext}
\begin{align}
\label{H3}
\hat{H}=&\underline{-\frac{\hbar^{2}\nabla_{R}^{2}}{2M}}-\frac{\hbar^{2}\nabla_{r}^{2}}{2\mu}+\frac{e^{2}B^{2}r^{2}}{8\mu c^{2}}(1-2\alpha\gamma +\alpha^{2}\xi_{3})+\frac{e\hbar}{Mc}(1-\alpha\gamma/2)[\mathbf{B}\times \mathbf{r}]\cdot (-i\nabla_{R})+\\ \notag &+\frac{e\hbar}{2\mu c}(\gamma - \alpha\xi_{2})[\mathbf{B}\times\mathbf{r}]\cdot (-i\nabla_{r}) + \underline{\alpha^{2}\frac{e^{2}B^{2}R^{2}}{8\mu c^{2}}} - \underline{\alpha\frac{e\hbar}{Mc}[\mathbf{B}\times \mathbf{R}]\cdot (-i\nabla_{R})}-\\ \notag &-\alpha\frac{e\hbar}{2\mu c}[\mathbf{B}\times\mathbf{R}]\cdot (-i\nabla_{r})-\alpha\frac{e^{2}}{4\mu c^2}(1-\alpha\gamma)[\mathbf{B}\times\mathbf{R}]\cdot[\mathbf{B}\times\mathbf{r}]-\frac{e^2}{\varepsilon\sqrt{r^2+L^2}},
\end{align}
\end{widetext}
where $\mu=m_{e}m_{h}/M$, $\gamma=(m_{h}-m_{e})/M$, $\xi_{3}=(m_{e}^{3}+m_{h}^{3})/M^{3}$, $\xi_{2}=(m_{e}^{2}+m_{h}^{2})/M^{2}$. Note, that in the case of a homogeneous magnetic field, when $\alpha=0$, the phase transformation (\ref{psi}) fully removes the action of the magnetic field on the CM of the exciton. However, in our case $B_{e}\neq B_{h},~\alpha\neq0$ and one sees that the magnetic field affects the motion of the center-of-mass. Moreover, strictly speaking it becomes impossible to separate relative and CM motions, and numerical treatment becomes necessary.

In Eq. (\ref{H3}) the first two terms correspond to the kinetic energy of the CM and relative motion of an exciton. The third term is simply a modified diamagnetic shift, the fourth term is a quasielectric field term which pushes the electron and the hole in opposite directions and the fifth term is a modified Zeeman-like term for the relative motion. All these terms exist for the homogeneous case ($\alpha=0$).\cite{Knox,Elliot,Hasegawa,Gorkov}

The new sixth, seventh, eighth and ninth terms (where $\alpha$ enters in the overall pre-factors) describe the effect of the magnetic field on the CM motion of an exciton. Finally, the tenth term represents the Coulomb attraction between the electron and the hole and is responsible for creation of the bound exciton state.

Let us now examine the Hamiltonian given by Eq. (\ref{H3}) in more detail. The terms corresponding to the exciton in a uniform magnetic field have been analyzed before. In small fields the binding energy of the exciton increases, while strong magnetic fields on the contrary lead to a transition to the magnetoexciton regime.\cite{Lozovik,Ruvinsky,MoskalenkoMagn} The crossover between these two regimes is well studied. In the present consideration we restrict ourself to the small magnetic field limit where cyclotron energy is much smaller than indirect exciton binding energy. As well, we do not consider the terms corresponding to the mixing of the center of mass and relative motion. The latter assumption can be justified by using the separate treatment of CM and relative motion of the exciton which corresponds Born-Oppenheimer approximation in atomic physics. Indeed, considering the relative motion as fast and CM motion as slow, and averaging over fast motion, it is easy to see that those terms cancel as $\langle\nabla_{r}\rangle=\langle[\mathbf{B}\times\mathbf{r}]\rangle=0$. Here we focus on additional terms which appear due to the gradient of the magnetic field in the $z$ direction.
\begin{figure}[t]
\includegraphics[width=1.0\linewidth]{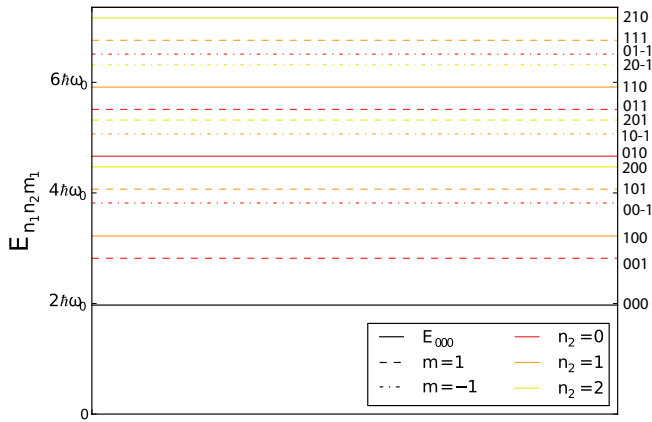}
\caption{(color online). The energy spectrum of the CM motion of a bound electron-hole pair in the inhomogeneous magnetic field given by Eq. (\ref{LL}).}
\label{Fig2}
\end{figure}

The part of the Hamiltonian corresponding to CM motion thus reads (only underlined terms in Eq. (\ref{H3}) are considered)
\begin{align}
\label{H5}
\hat{H}&=\frac{\hbar^{2}}{2m_{e}}(-i\beta_{e}\nabla_{R}-\alpha\frac{e}{\hbar c}\mathbf{A}(\mathbf{R}))^{2}+\\&+\frac{\hbar^{2}}{2m_{h}}(-i\beta_{h}\nabla_{R}-\alpha\frac{e}{\hbar c}\mathbf{A}(\mathbf{R}))^{2}. \notag
\end{align}
which can be simplified to
\begin{align}
\label{H6}
\hat{H}&=\frac{1}{2M}\Big(\hat{\mathbf{P}}-\frac{e^{*}}{c}\mathbf{A}(\mathbf{R})\Big)^{2}+\frac{e^{*2}B^2R^2\gamma^{2}}{8c^{2}\mu},
\end{align}
where $\hat{P}=-i\hbar\nabla_R$ is the CM momentum of the electron-hole pair and $\mathbf{A}(\mathbf{R})=[\mathbf{B}\times \mathbf{R}]/2$. One can see that expression (\ref{H6}) looks exactly the same as the Hamiltonian of a 2D particle with effective charge $e^*$ placed in a magnetic field perpendicular to the plain in the presence of an in-plane external parabolic confining potential $U=e^{*2}B^ 2R^2\gamma^{2}/8c^{2}\mu$. The value of the effective charge of the indirect exciton which describes its response to an inhomogeneous magnetic field is
\begin{equation}
e^{*}=2 \alpha e = \frac{\Delta B}{B}e.
\label{e_eff}
\end{equation}
As one can expect, it is proportional to the difference of the magnetic field acting on the electron and hole subsystems and vanishes for the uniform case. Note that this effective charge describes only the response to the magnetic field and an external electric field will still not affect the motion of an exciton which remains a neutral particle.
\begin{figure}[t]
\includegraphics[width=0.9\linewidth]{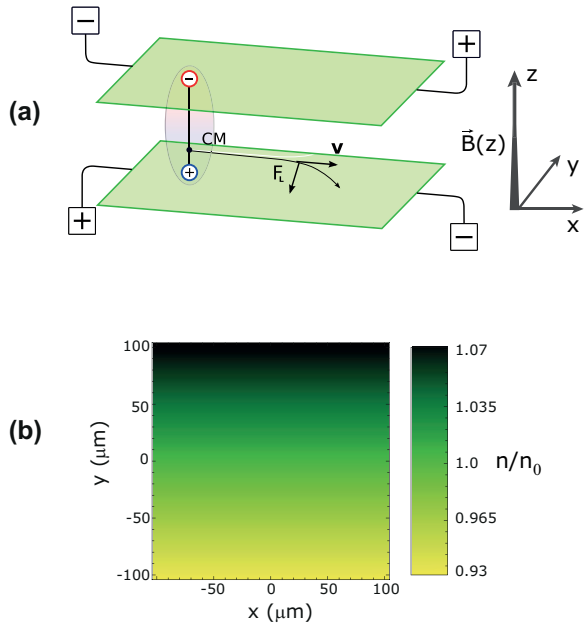}
\caption{(color online). (a) Geometry of the counterflow experiment. Applying the voltage of different polarities for the electron and hole subsystems leads to the generation of the flux of indirect excitons in $x$ direction. (b) The profile of 2D density of indirect excitons in the inhomogeneous magnetic field. The drift velocity is taken $V=10^4$ cm/s, magnetic field difference $\Delta B=1$ mT and background concentration $n_{0}=10^{9}~cm^{-2}$. The Hall effect causes deviation of exciton density equal to $7 \%$.}
\label{Fig3}
\end{figure}

The solution of the eigenvalue problem for a particle in a magnetic field in the presence of parabolic confined potential is well known.\cite{Galitskii} This allows us to write the expression for the energy spectrum of an indirect exciton in an inhomogeneous magnetic field as

\begin{eqnarray}
\label{LL}
E_{n_{1}n_{2}m_{1}}=\hbar\sqrt{\widetilde{\omega}^{2}+\omega_{0}^{2}}\Big(n_{1}+\frac{1}{2}+\frac{|m_{1}|}{2}\Big)+\\
 \nonumber+\frac{\hbar \omega_{0}}{2}\Big(n_{2}+\frac{1}{2}\Big)-\frac{\hbar \omega_{0}}{2}m_{1},
\end{eqnarray}
where $n_{1}$, $n_{2}$ and $m_{1}$ are integer numbers. The quantum number $m_1$ corresponds to an angular momentum. The cyclotron frequency $\omega_{0}$ and the potential well quantization frequency $\widetilde{\omega}$ read
\begin{equation}
\omega_{0}=\frac{e^{*}B}{Mc}=\frac{e \cdot \Delta B}{Mc}=\frac{\mu}{m_{h}-m_{e}}\widetilde{\omega}.
\label{omega}
\end{equation}
Energy levels given by Eq. (\ref{LL}) are shown in Fig. \ref{Fig2}. Note, that differently from the case of the Landau quantization for the free electrons, the spectrum is not equidistant. It should also be noted that the energy distance between the levels corresponding to the quantization of CM motion of an exciton is much smaller than the distance between excitonic levels appearing due to the quantization of relative motion.  The spectrum in Fig. \ref{Fig2} thus can be interpreted as ``fine structure'' of $1s$ exciton state in an inhomogeneous magnetic field.

For $\Delta B=1$ mT we can estimate $\hbar\omega_0=0.2~\mu$eV, which is much smaller than the thermal energy corresponding to $1~K$. Therefore, direct observation of quantization of the energy levels is hardly possible. One thus needs to propose another way of the observation of the effect of an inhomogeneous magnetic field on indirect excitons. Note that the binding energy of the exciton in typical GaAs/AlGaAs structure with $L=12$ nm separation between centers of quantum wells is $E_{b}=3.67$ meV, which is much larger than $\hbar\omega_0$. This confirms the validity of the Born-Oppenheimer approximation which we used.

We propose an experiment where the Lorentz force for neutral excitons can be detected analogously to the classical Hall effect case (see Fig. \ref{Fig3}). The key point is generation of a flow of excitons, which is a more complicated task than for the case of electrons since excitons are electrically neutral particles and application of an electric field does not lead to the appearance of the exciton current. However, this current can be created in a counter-flow experiment, where polarity of the drain-source voltage is different for the layers of the electrons and the holes. The counterflow techniques is a well developed method and was widely used by Eisenstein and co-workers for the study of the physical properties of Quantum Hall bilayers.\cite{Eisenstein}

The drift velocity of an indirect exciton in a counter-flow experiment will be governed by the value of the electric fields in electron and hole layers, $E_{e,h}$, and mobilities of the electrons and the holes $\mu_{e,h}$:

\begin{equation}
\mathbf{V}=\frac{\mu_{e}\mu_{h}}{\mu_{e}+\mu_{h}}(\mathbf{E}_{h}-\mathbf{E}_{e}).
\end{equation}
One should note that the velocity can not be too large since the magnetic field tends to unbind fast moving excitons.\cite{Knox,Lozovik}
\begin{figure}[t]
\includegraphics[width=0.9\linewidth]{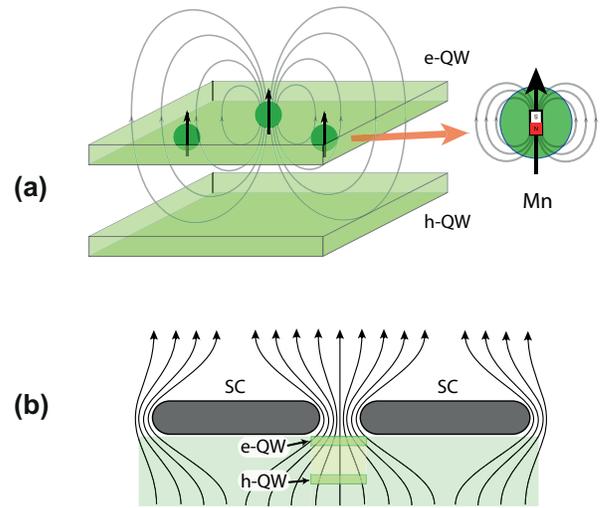}
\caption{(color online). Sketches of the setup with inhomogeneous magnetic field created by ferromagnetic doping in one QW (a) and due to the Meissner effect induced by superconducting film (b).}
\label{Fig4}
\end{figure}

Due to non-zero effective charge of the indirect exciton the Lorentz force acting on CM coordinates appears $\mathbf{F}_L=e^{*}\mathbf{V}\times \mathbf{B}/c$. It deviates the moving exciton in the direction perpendicular to the direction of the electric field $(\mathbf{E}_{h}-\mathbf{E}_{e})$. In the stationary regime this force should be compensated by the force provided by a gradient of potential energy of the exciton-exciton repulsion, which can be estimated as $\mathbf{F}_{p}=-U_{0}\nabla n(\mathbf{r})$, with $U_0\approx 4\pi e^2L/\varepsilon$,\cite{Kyriienko} where $L$ is the distance between the QWs with the electrons and holes and $\varepsilon$ is a dielectric permittivity.  Putting $\mathbf{F}_L=\mathbf{F}_{p}$ one can determine the spatial profile of the exciton concentration in the Hall bar geometry shown in Fig. \ref{Fig3}, where the flux of the excitons is directed along the $x$ axis,
\begin{equation}
n(y)=n_{0}+\frac{e V \Delta B}{U_{0}}\frac{y}{2},
\label{ny}
\end{equation}
and $n_0$ denotes a background 2D density of the indirect excitons.  The density profile is shown in Fig. \ref{Fig3} for the magnetic field $\Delta B=1~mT$, $n_{0}=10^9~cm^{-2}$ and drift velocity $V=10^4~cm/s$. The variation of the density can be measured by analyzing the spatial variation of the blueshift of the exciton photoluminescence in the near-field experiment. One should also note that for the described Hall effect for exciton the relevant quantity is difference of magnetic fields $\Delta B$ and not an absolute value $B$, which can be chosen arbitrarily small.

Previously we considered an abstract situation of inhomogeneous magnetic field. The realization of such situation in experiment is a non-trivial task. For an infinite sample the Gauss's law for magnetism forbids the possibility of magnetic field generation with gradient in growth direction. However, the finiteness of studied structure allows one to consider a situation where magnetic field lines are non-uniform in $xy$ plane, but characteristic magnetic length is much larger than size of the sample. Thus, a magnitude of magnetic field is locally homogeneous and different for electron and hole layer. These can be realized in practice using several schemes.\cite{Freire} 

First, the difference in the magnetic fields acting on electrons and holes $\Delta B$ can be achieved \textit{e.g.} by using magnetic semiconductor material for one of the QWs (Fig. \ref{Fig4} (a)). In particular, the inhomogeneity can be realized by growing a Ga$_{1-x}$Mn$_x$As quantum well for electrons, where $x$ denotes a fraction of manganese magnetic impurities. Experimentally, $x$ usually lies in the range from $0.01$ to $0.05$,\cite{Munekata,Jungwirth} but in state of art samples it can achieve $0.08$ fraction of manganese.\cite{Ohno}
The magnetic field induced by a single Mn atom as a function of distance from it $r$ reads as
\begin{align}\label{MnB}
B_{z}(r)=\frac{\mu_0}{4\pi}\frac{2M_0}{(a_0^2+r^2)^{3/2}},
\end{align}
where $\mu_0$ is a vacuum permeability, $a_0$ is a radius of the atom and $M_0$ denotes its magnetic moment. For Mn atom it is equal to $M_0=8\mu_B$, where $\mu_B$ is the Bohr magneton.\cite{Munekata,Jungwirth,Tandon}

Using Eq. (\ref{MnB}) we can estimate that an electron moving through the structure with $x=0.05$ will experience a magnetic field of magnitude $B=0.2$ mT. The magnetic field in the hole QW, which is separated from electron QW by distance $L=12$ nm, is however three orders of magnitude smaller, and total magnetic field difference is $\Delta B\approx 0.2$ mT. If we take $x$ to be $0.08$ the magnetic field can exceed $\Delta B \approx 1$ mT. Therefore the values of $\Delta B$ used for calculation of Hall effect in the article can be in principal realized in experiments.

Second, an inhomogeneous field can be generated using the Meissner effect. In the vicinity of superconducting film (SC) the magnetic field lines are strongly non-uniform and this can be used for local generation of magnetic field gradient in growth direction (see sketch in Fig. \ref{Fig4} (b)). 
 
Finally, another possible way of non-uniform magnetic field generation relies on the quadrupole magnet field. In the volume between four magnets magnetic field lines are strongly non-uniform and magnetic field magnitude changes rapidly in the plane of the system.\\

\section{Conclusions} In conclusion, we proposed a method of generation of the effective magnetic field acting on a system of electrically neutral spatially indirect excitons. It appears in the case of applied weak magnetic field which is inhomogeneous in $z$ direction. We analyzed the energy spectrum of the system and predicted the existence of an analog of the classical Hall effect for spatially indirect excitons. It can be detected in a counter-flow experiment where asymmetric photoluminescence profile of indirect excitons is expected.

The work was partially supported by FP7 IRSES ``SPINMET'' project. O. Kyriienko acknowledges the support from the Eimskip Fund.

%%% References %%%

\end{document}